\documentclass[11pt, letterpaper]{article}

\usepackage[margin=1in]{geometry}
\usepackage{amsmath}
\usepackage{graphicx}
\usepackage{times}
\usepackage{tabularx,booktabs}
\usepackage{authblk}

\author[1]{Puqi Zhou}
\author[1]{Charles R. Twardy}
\author[1]{Cynthia Lum}
\author[1]{Myeong Lee}
\author[1]{David J. Porfirio}
\author[1]{\\Michael R. Hieb}
\author[2]{Chris Thomas}
\author[1]{Xuesu Xiao}
\author[1]{Sungsoo Ray Hong}

\affil[1]{George Mason University, USA}
\affil[2]{Virginia Tech, USA}
\date{}

\title{\textbf{Applying Ground Robot Fleets in Urban Search: Understanding Professionals’ Operational Challenges and Design Opportunities}}

\begin{document}

\maketitle

\section*{Abstract}

Urban searches demand rapid, defensible decisions and sustained physical effort under high cognitive and situational load. Incident commanders must plan, coordinate, and document time-critical operations, while field searchers execute evolving tasks in uncertain environments.
With recent advances in technology, ground-robot fleets, paired with computer-vision-based situational awareness and LLM-powered interfaces offer the potential to ease these operational burdens.
However, no dedicated studies have examined how public safety professionals perceive such technologies or envision their integration into existing practices, risking building technically sophisticated yet impractical solutions.
To address this gap, we conducted focus-group sessions with eight police officers across five local departments in Virginia.
Our findings show that ground robots could reduce professionals’ reliance on paper references, mental calculations, and ad-hoc coordination,  alleviating cognitive and physical strain in four key challenge areas: (1) partitioning the workforce across multiple search hypotheses, (2) retaining group awareness and situational awareness, (3) building route planning that fits to the lost person profile, and (4) managing cognitive and physical fatigue under uncertainty.
We further identify four design opportunities and requirements for future ground-robot fleet integration in public-safety operations: (1) scalable multi-robot planning and control interfaces, (2) agency-specific route optimization, (3) real-time replanning informed by debrief updates, and (4) vision-assisted cueing that preserves operational trust while reducing cognitive workload.
We conclude with design implications for deployable, accountable, and human-centered urban-search support systems.

\section{Introduction}
Urban search operations demand considerable mental and physical effort from public safety professionals (professionals, hereinafter).
The life-or-death nature of these missions—where every minute matters—further intensifies the challenge for local law enforcement to conduct timely and defensible searches~\cite{adams2007search}.
Professionals operate under severe time pressure and uncertainty, resulting in significant cognitive load~\cite{feinberg2021reconceptualizing}.
Despite technological progress, prior studies show that urban searches remain largely manual, with limited integration of advanced systems~\cite{ferguson2023training}.
Practitioners still rely on reference materials (e.g., Lost Person Behavior~\cite{koester2008lost}), field manuals~\cite{pub1991national}, and heuristically derived procedures~\cite{wollan2004incorporating}.
Incident commanders must synthesize lost person profiles, coordinate multiple teams, update plans rapidly, and maintain shared situational awareness~\cite{young2007urban}.
Meanwhile, field searchers face substantial physical strain as they navigate complex terrains and adapt to evolving conditions~\cite{vinnikov2022occupational}.
While practitioners’ expertise remains indispensable, the continued reliance on manual coordination limits scalability and responsiveness.
Given rapid advances in sensing, robotics, and AI technologies, it is crucial to explore how these tools can effectively augment human effort, improving the efficiency, accuracy, and defensibility of future urban search operations.


Deploying ground robots in urban search operations can enhance professionals’ situational awareness and reduce both physical and cognitive demands.
Recent advances have made it feasible to employ autonomous ground-robot fleets as force multipliers~\cite{Roberts2016UnmannedVC}, extending operational reach, providing persistent ground-level sensing, and complementing human searchers~\cite{Grayson2014SearchR}.
Modern ground robots now exhibit reliable mobility across paved and semi-structured terrains~\cite{xu2024reinforcement}, streaming stabilized, eye-level video that captures detailed facial, gait, and contextual cues often unavailable from aerial platforms~\cite{alotaibi2019lsar}.
These capabilities enable incident commanders to coordinate robot fleets at scale and maintain continuous situational coverage.
In parallel, advances in computer vision have enabled robust object detection~\cite{du2022unknown}, person retrieval~\cite{jiang2023cross}, and scene understanding~\cite{zhou2024hugs} in challenging urban conditions.
Large language models (LLM) further augment these capabilities by providing intelligent interfaces that connect professionals and robot fleets—supporting route planning~\cite{meng2024llm}, information synthesis, and shared situational awareness across teams~\cite{cao2025exploring}.
Recent LLM-based systems demonstrate the ability to reason over spatial-temporal context for urban location-based services~\cite{jiang2024urbanllm} and to generate contextually rich, adaptive scenarios for law-enforcement training~\cite{violakis2025leveraging}.

While recent technological progress has made ground robots increasingly capable, adopting them into public safety operations remains highly challenging, as real-world deployment is fraught with uncertainty, context-specific constraints, and organizational complexity.
Law-enforcement workflows are highly contextualized, agency-specific, and governed by strict protocols.
Recent deployments of ground robots, such as the NYPD’s use of the K5 robot~\cite{nytimes2023K5}, illustrate these challenges: despite its technical sophistication, the robot faced public skepticism, operational constraints, and limited field utility.
Such examples underscore that without a deep understanding of existing workflows, information requirements, and organizational factors, we risk developing technically advanced systems that remain incompatible with operational realities.
Prior research has focused primarily on robotic performance metrics, such as navigation accuracy~\cite{xiao2022motion, 11247618} and coordination efficiency~\cite{rekleitis2002multi, burgard2005coordinated, 11247499}, with limited investigation of practitioners’ workflows, decision-making, and organizational constraints.
Burke et al.~\cite{burke2004moonlight} observed that professionals often spent disproportionate time reconstructing robot state and spatial context, struggling to align the robot’s camera view with their mental model during disaster response drills. This highlights the enduring challenge of human–robot coordination in high-stakes, time-sensitive environments.

We conducted a study with public safety professionals to examine their current urban search practices, perceived challenges, and envisioned opportunities for adopting ground robotics, focusing on how they anticipate these systems could support or complicate coordination, public engagement, and operational decision-making in urban search contexts.
Our study involved eight police professionals from five local law enforcement agencies in Virginia, who participated in focus group discussions centered on their experiences and perspectives related to technology adoption in search operations.
Through qualitative analysis, we first identified the current urban practice and high-priority challenges that professionals encounter, along with their views on where ground robotics could provide meaningful operational support and why such support matters.
Next, we then derived implications for design toward developing human–robot collaborative systems that can meaningfully augment search operations, translating practitioners’ front-line expertise into concrete design implications for deployable, accountable, and trustworthy search-support technologies.
Finally, we distilled a set of design requirements that future systems should embody to ensure practicality and relevance across three technological domains: human–computer interaction, robotics, and computer vision.

This work contributes an in-depth understanding of how highly skilled public safety professionals perceive the adoption of ground-robot fleets in practice, offering insights that can guide researchers across HCI, robotics, and computer vision in developing practically useful and adoptable tools for future public-safety operations.

\section{Literature Review}
Understanding how ground robot fleets can support urban lost-person searches is an emerging problem space. 
While few studies address this question directly, related work in search practices and decision support, robotics in search and rescue, and AI-assisted public safety offers useful foundations.
We highlight both progress in each domain and remaining gaps that motivate our empirical investigation.

\subsection{Search Operation and Decision Making}
The inherent uncertainty and dynamic nature of lost person behavior, shaped by individual psychology, environmental factors, and evolving circumstances, make location forecasting fundamentally challenging for both technological systems and human professionals.
This difficulty is compounded by the heavy cognitive and physical demands that lost person searches impose on professionals~\cite{young2007urban, prati2010self}.
The complexity of these operations has driven researchers and professionals to develop complementary approaches that improve search efficiency and enable high-probability search strategies.
Contemporary search planning draws from multiple analytical traditions: classical search theory~\cite{stone2007theory} and information path planning algorithms~\cite{hashimoto2022agent, nguyen2023finding}. 
The International Search and Rescue Incident Database (ISRID)~\cite{koester2010international} has emerged as a critical resource, offering statistical distributions of travel distances and behavioral patterns across different demographics and scenarios~\cite{koester2008lost}. 
Building on these foundations, computational models now incorporate probabilistic reasoning to predict likely lost person locations and movement routes through ring model zoning~\cite{koester2008lost, sava2016evaluating}, agent-based walking simulations~\cite{hashimoto2022agent, ewers2025predictive}, and terrain-informed heatmaps that integrate weather conditions and person profiles~\cite{ewers2023optimal}.
However, most research focuses on prediction models for specific person categories (e.g., hiker~\cite{doherty2014analysis}), with limited exploration of how these models can generalize across diverse 45 person types~\cite{koester2008lost} or incorporate contextually customized information unique to individual cases.

In practice, searches rarely proceed under a single ``optimal'' decision-making as computational tools that propose a singular solution.
Field teams plan and replan in cycles, updating tactics as conditions and information evolve. 
Incident commanders rely on field manuals, acquired experience, and heuristics—often satisfied by choosing timely, workable options over exhaustive optimization~\cite{cohen2015investigation, hill2012cognition}.
Drawing on possible patterns and actuarial cues of the lost person, experts envision plausible, executable courses of action~\cite{ koester2008lost}.
Coordination is typically manual and consensual within small teams, with priorities revised as reports, clues, and eyewitness accounts arrive~\cite{ferguson2023training, young2007urban}. 
The shared information updating across people and externalized in maps, radio traffic, and emerging digital tools, with decision authority enacted collectively rather than vested in a single commander~\cite {alharthi2021activity}. 
This mismatch between algorithms pursuing single optima versus professionals orchestrating collaborative, adaptive strategies establishes a clear design mandate: robotic systems must augment, not disrupt, these distributed, update-driven workflows.

\subsection{Robotics in Search and Rescue}
The deployment of robotic systems in search and rescue operations has evolved significantly over the past two decades, driven by advances in autonomy, navigation, sensing, and coordination algorithms, yet these systems have primarily targeted post-disaster scenarios and limited ground robot integration. 
Following the 2001 World Trade Center collapse, ground robots were first deployed for structural inspection and victim search~\cite{ko2009intelligent, casper2003human}. 
These early uses demonstrated robots’ potential to access hazardous or confined spaces but also revealed challenges in maintaining situational awareness and interpreting sensor data under pressure~\cite{burke2004moonlight, casper2003human}. 
Subsequent work in navigation, mapping, and human–robot interaction sought to address these limitations by improving autonomy and information presentation.
Unmanned Aerial Vehicles (UAVs) have been widely explored for wilderness and post-disaster search due to their ability to rapidly survey large and inaccessible areas~\cite{doroodgar2014learning}. 
However, UAVs face critical limitations in urban environments, including regulatory restrictions, short flight times, weather sensitivity, and occlusions from dense infrastructure~\cite{glantz2020uav, quero2025unmanned}. 
Their top-down imagery also lacks the pedestrian-level detail—such as facial features and gait patterns—necessary for identifying lost persons.  
By contrast, Unmanned Ground Vehicles (UGVs) provide persistent, close-range sensing on paved and semi-structured terrain~\cite{xiao2022motion}. 
Modern wheeled platforms integrate stabilized cameras, LiDAR, and thermal sensors to detect subtle environmental cues that might escape human notice~\cite{hong2019investigating, kashino2020aerial}.
Yet, despite improved autonomy and significant light operator supervision, the current UGVs' integration into dynamic, multi-team search workflows remains limited.

Multi-robot systems offer clear advantages for search operations through parallelization, redundancy, and complementary sensing, yet their deployment in urban search remains limited by organizational constraints, legal responsibilities, and the need for explainable and auditable plans aligned with agency protocols.
Recent coordination research has focused on coverage, task allocation, and cooperative exploration~\cite{queralta2020collaborative}, optimizing efficiency and energy use under simplified conditions while overlooking command structures, evolving priorities, and operator workload.
The cognitive demands of monitoring multiple robots often offset these efficiency gains, underscoring the need for human-centered interfaces that translate algorithmic outputs into actionable, interpretable plans~\cite{williams2020collaborative}.
Moreover, embedding human expertise—such as reasoning about routes, terrain, and behavioral cues—into robotic decision frameworks remains an open challenge~\cite{francos2023role}.
Although early work, such as Ferguson et al.~\cite{ferguson2023training}, has begun to incorporate expert knowledge into coordination logic, how multi-robot systems can be effectively integrated into real urban public-safety workflows remains an unresolved problem.

\subsection{AI-assisted Public Safety}
Artificial intelligence technologies are increasingly being explored to augment human capabilities in public safety operations, offering new possibilities for information processing, decision support, and task automation, yet integrating robot-captured video and contextual data with geographic route planning has received limited attention.
These technologies span computer vision, natural language processing, and interface design, each contributing distinct capabilities to enhance operational effectiveness.
Computer vision techniques have shown promise in enhancing situational awareness during emergency operations, enabling onboard detection, tracking (including re-identification), and scene understanding from both stationary sensors and robot-mounted cameras operating in dynamic environments~\cite{dhiman2022firefighting}.
For example, UAV-based thermal imaging and multi-spectral sensing enhance detection of persons or vehicles across challenging conditions~\cite{barisic2025unlocking}.
Yet most models are trained on fixed surveillance footage and struggle with the motion, variable lighting~\cite{8970561, guo2020motion_blur_dnn_neurips}, and computational limits of mobile ground platforms. 
These computer vision capabilities still enable automated analysis of imagery and video feeds, reducing the cognitive burden on human operators who would otherwise need to review extensive footage manually.

Large language models (LLM) represent an emerging technology for contextual training, analysis, and decision support in public safety operations~\cite{chen2024ekell}. 
Recent studies explore LLMs for urban computing, location-based services~\cite{xiao2025llmadvisorllmbenchmarkcostefficient, li2024llm_poi}, and incident summarization. 
The ability of LLMs to understand context and reason about textual information may prove valuable for supporting incident commanders in information-rich environments~\cite{otal2024crisis}. 
Law enforcement professionals show the need for domain-specific explanations~\cite{herrewijnen2024xailaw}, clear rationales for recommendations~\cite{bucinca2021trust}, and mechanisms to challenge AI outputs~\cite{karusala2024publicservices}.
However, adopting such systems in high-stakes domains raises known challenges: over-reliance, opaque reasoning, and difficulties in verifying recommendations.
For operational deployment, outputs must remain verifiable, policy-aligned, and presented in ways that support, not replace professional judgment.
The effectiveness of AI-assisted systems ultimately depends on interface designs that support appropriate reliance and effective collaboration. 
Studies of transparent-display detail show a speed–accuracy tradeoff—more detail can boost accuracy but increase cognitive load and slow responses~\cite{spatharioti2025llmsearch}.

Despite progress across these domains, significant gaps remain in understanding how ground robot fleets can effectively support urban lost-person searches. 
While computational models offer sophisticated prediction capabilities, they inadequately account for the distributed, collaborative nature of actual search operations. 
Robotic systems demonstrate technical capabilities but lack clear pathways for integrating domain expertise into operational deployments. 
AI technologies show promise for information processing but require careful interface design to support effective human-robot collaboration.
Critically, existing research has primarily evaluated robotic search systems in controlled or simulated environments, with limited investigation of how these systems would be adopted and utilized by actual search professionals in realistic operational contexts. 
This gap between technical capability and operational utility motivates our empirical investigation into how experienced search professionals envision deploying and coordinating ground robot fleets for urban search scenarios.
\section{Method}
Our goal is to understand how urban public safety professionals currently conduct lost-person searches and how affordable ground robot fleets could be integrated into these workflows in realistic agency settings, including smaller departments with limited resources. 
We conducted open-ended conversational focus group sessions with police professionals who have direct experience in urban search.
While ethnographic methods like participant observation could also support our goals, given that ground robot fleets have not yet been widely deployed in urban search operations, we opted for focused group sessions as they allowed us to discuss and explore hypothetical integration scenarios.
We describe the methodology we followed to recruit participants, conduct sessions, and analyze the collected data.

Specifically, our work is driven by three research questions:

\noindent\textbf{RQ1:} How do urban public safety professionals currently plan, execute, and document lost-person searches, and what practical constraints shape their strategies?

\noindent\textbf{RQ2:} Where do professionals see meaningful roles and clear limits for autonomous ground robot fleets within real-world urban search operations?

\noindent\textbf{RQ3:} What design opportunities emerge for human-centered tools that integrate ground robots, computer vision, and AI-assisted reasoning into accountable, professional-led urban search workflows?

\subsection{Participants}
Police professionals have demanding workloads and are bound by strict confidentiality protocols, making direct recruitment for research studies challenging.
To recruit professionals, we used a convenience sampling strategy to identify and connect police professionals with urban search experience.
Three authors visited three local police departments with six professionals in Virginia and invited two professionals from different departments to participate in focus group sessions at our university.
All participants (see Table~\ref{tab:participant-profiles}) had direct experience with lost-person search operations, ranging from detectives who assist with searches to senior professionals who have served as incident commanders.
\begin{table*}[t]
\centering
\vspace{-1.5em}
\begin{tabularx}{\textwidth}{l l X c}
\toprule
\textbf{PID} & \textbf{Rank} & \textbf{Agency} & \textbf{Years Exp.} \\
\midrule
P1 & \small{Lieutenant} & \small{Arlington County Police Department} & 23 \\
P2 & \small{Detective}  & \small{Manassas City Police Department} & 9 \\
P3 & \small{Detective}  & \small{Loudoun County Sheriff's Office Administration} & --- \\
P4 & \small{Sergeant}        & \small{Loudoun County Sheriff's Office Administration} & --- \\
P5 & \small{Captain}    & \small{Fairfax City Police Department} & 19 \\
P6 & \small{Detective}  & \small{Fairfax City Police Department} & 25 \\
P7 & \small{Detective}  & \small{Fairfax City Police Department} & 5 \\
P8 & \small{Major}    & \small{Fairfax County Police Department} & 16 \\
\bottomrule
\end{tabularx}
\caption{Participant Profiles.}
\vspace{-1.5em}
\label{tab:participant-profiles}
\end{table*}

\subsection{Study Procedure}
Each focus group session lasted approximately 90 minutes and was conducted in a private conference room.
The session was designed to elicit both current practice and informed reflections on the potential use of ground robot fleets.
At the beginning of the session, the corresponding author introduced the study goals and explained that our focus was on affordable, ground-based robotic platforms that could realistically be acquired and operated by small and mid-sized agencies, rather than specialized, high-cost systems.
The corresponding author then presented a brief slide deck showcasing an example of applying affordable ground robot fleets in campus patrol with police-centered video sensemaking technologies and richful information user interface.
The presentation was intentionally constrained to avoid overselling capabilities and to leave room for participants to particulate concerns, constraints, and alternative uses.
This framing helped ground discussions in realistic deployment scenarios rather than speculative future technologies.

Following the introduction, the three authors facilitated open-ended discussions organized around four main themes: (1) Background and role, (2) Current urban search workflows, (3) Operational challenges, (4) Ground robot integration.
Throughout the sessions, we used probing to elicit concrete examples and operational detail. 
For instance, when participants mentioned challenges with ``coordination'' we asked them to walk us through a specific search scenario step-by-step, describing what information they needed, what decisions they made, and what communication occurred.
We also used scenario-based probes: ``Imagine you have three ground robots available for a lost person search in a residential area. How would you deploy them? What would you need to know to decide their routes?''

Rather than audio recording, we took detailed written notes to respect participants’ confidentiality concerns and departmental protocols. Three researchers were present at each focus group session: one facilitated the discussion while the other two independently documented participants’ statements, concrete examples, and operational constraints. Immediately following each session, the three researchers met to consolidate notes, reconcile discrepancies, and elaborate key points while details were fresh.
To strengthen analytic trustworthiness given the absence of verbatim transcripts, we retained both the independently taken notes and the consolidated version to preserve an audit trail from raw capture to synthesized records; documented consolidation decisions when reconciling discrepancies; and used investigator triangulation during analysis, with multiple authors iteratively reviewing interpretations and challenging emerging claims against the consolidated records. We also prioritized capturing participants’ language for critical claims (e.g., constraints, failure modes, and decision rationales) to reduce over-interpretation.
During sessions, the facilitator periodically summarized key points back to the group to confirm shared understanding and correct misunderstandings in real time.

We analyzed the consolidated focus group notes using an iterative, collaborative thematic analysis process. After each session, the first author compiled and organized the consolidated notes and conducted an initial round of inductive coding focused on workflow steps and information artifacts, constraints and breakdowns, coordination and documentation practices, and perceived roles/limits of envisioned robot fleets.
The first author then clustered related codes into candidate themes aligned with our three research questions.
The other two authors independently reviewed the coded notes and candidate themes, checking them against the consolidated records, surfacing alternative interpretations, and identifying missing or weakly supported claims. Through regular analytic meetings, we iteratively refined code definitions and theme boundaries, resolving disagreements through discussion and returning to the notes for adjudication. This process converged on the four challenges and four design opportunities reported in Section 4, each grounded in recurring patterns across participants and supported by concrete operational examples.
\section{Results}
Our analysis of focus group discussions with eight police professionals from five Virginia law enforcement agencies reveals how practitioners currently conduct urban search operations, the critical challenges they face, and their envisioned opportunity for ground robots. 
\begin{table*}[t]
\centering
\vspace{-1.5em}
\begin{tabularx}{\textwidth}{l X c c c}
\toprule
\textbf{Agency (PD/SO)} & \textbf{Jurisdiction} &
\textbf{Pop. (2024)} & \textbf{Land (sq mi)} & \textbf{Sworn} \\
\midrule
\small{Arlington County Police Department} &
\small{Arlington County, VA} &
\small{239,807} & \small{26.00} & \small{377} \\

\small{Manassas City Police Department} &
\small{City of Manassas, VA} &
\small{43,616} & \small{9.84} & \small{99} \\

\small{Loudoun County Sheriff's Office} &
\small{Loudoun County, VA} &
\small{443,380} & \small{515.74} & \small{$\sim$700} \\

\small{Fairfax City Police Department} &
\small{City of Fairfax, VA} &
\small{26,340} & \small{6.24} & \small{69} \\

\small{Fairfax County Police Department} &
\small{Fairfax County, VA} &
\small{1,160,925} & \small{391.02} & \small{$\sim$1,500} \\
\bottomrule
\end{tabularx}
\caption{Agency overview for participants' departments.}
\vspace{-1.5em}
\label{tab:agency-overview}
\end{table*}

\subsection{Current Urban Search Practices}
\noindent\textbf{Profile-Driven Search Planning}
Across agencies, professionals described urban searches as fundamentally profile-driven: they tailor search plans to the lost person’s type and, more importantly, to that individual’s profile, including typical behavior, time of disappearance, and the environment from which they went missing. 
professionals reported that a small set of recurring profiles: children, elderly people with dementia, and individuals with mental health conditions, dominates urban cases (P1–P8). 
For each profile, they rely on guidance from \emph{Lost Person Behavior}~\cite{koester2008lost} and empirically grounded expectations about where the person is likely to move or stay. 
For example, dementia cases are often conceptualized as “hiders” (e.g., backyards, behind houses, inside small rooms) (P1–P6) or “walkers” who remain on or near sidewalks, with P1 estimating that roughly half of older adults are ultimately found at or along sidewalks. 
Autistic or non-verbal individuals are expected to cycle through a narrow set of familiar points of interest (PoIs), while children are assumed to operate within small, familiar mobility maps around home or known landmarks and to be strongly drawn to water, which professionals frame as points of danger (PoDs) (P1, P2).
These category-specific patterns are further refined by time-of-day expectations: participants consistently associated dementia- and mental-health–related cases with nighttime disappearances, whereas child cases were more often anticipated during the day, particularly from afternoon into dusk (P1–P4, P6). 
Together, these rules-of-thumb, combined with formal guidelines~\cite{koester2008lost, young2007urban}, provide a structured, experience-based starting point for drawing initial search areas and dispatching units.

At the same time, professionals stressed that these templates are immediately customized with person-specific and environment-specific information. They routinely assemble a ranked list of the individual’s meaningful places—sites they frequently visit, talk about, or were previously found—which are treated as high-priority PoIs and cross-checked against formal lost-person behavior guidance (P1, P6). 
professionals also reason about how the originating environment (e.g., nursing home, dense residential block, downtown event) shapes plausible routes and hiding opportunities, noting that the “same” profile can behave differently across these contexts (P6). 
Appearance information is folded into this planning but remains fragile: participants reported that they can obtain a photograph in most cases, often supplemented by surveillance or Ring camera footage that provides reliable clothing and posture details (P1, P4). 
However, they also described systematic noise in verbal clothing descriptions—especially for color terms such as green versus blue or purple versus red—which limits their reliance on witness-reported colors (P1, P2). 
Overall, our findings show that “profile-driven search” in urban policing is a complex, cognitively demanding operational practice that combines profile-specific heuristics, temporal and environmental expectations, and curated lists of points of interest to decide where, when, and how to search.

\noindent\textbf{Coordination in Planning}
Across agencies, participants described search planning as a highly coordinated, time-sensitive process in which incident commanders must synthesize fragmented information while mobilizing limited personnel. 
Patrol professionals begin by completing the Virginia standard lost person form to establish an initial profile (P1, P4, P7), after which detectives typically have up to two hours to deepen that profile by gathering historical information and verifying critical details (P7). 
During this period, incident commanders work to assemble sufficient manpower, yet participants noted that staffing constraints often make it difficult to secure the number of professionals they believe the case requires (P1,P6); only officially designated ``critical missing'' cases allow them to raw on additional resources across units or agencies (P5, P6). 
Once a lost person profile is compiled, commanders translate it into a coordinated search plan that distributes responsibility across patrol professionals, detectives, and, when available, SAR teams. 
They described three interlocking planning logics—\emph{point-based} (P1, P2), \emph{area-based}(P3, P4), and \emph{linear-feature–based}(P6). 
Depending on the case, commanders may prioritize familiar points of interest (PoIs), points of danger (PoDs) such as water sources, and recently visited locations (P1–P3), or they may organize searches along linear features like sidewalks, creeks, and road corridors (P6).
In routine police practice, professionals emphasized, they rarely have detailed predefined routes; instead, they start from the last known point (LKP), known direction of travel, PoI history, and environmental cues to carve the map into zones.
These zones are then assigned to teams (P3, P4), with team leaders further dividing them into directions or small sectors for individual searchers, adjusting whether searches are conducted by teams or solo professionals based on available personnel.

Coordination continues throughout the operation as teams report back clues, sightings, and negative search results. 
Participants explained that search plans are repeatedly updated based on this incoming information and on short team briefings and debriefings (P3, P6). 
Some agencies explicitly structure this process using a ring-model approach derived from \emph{Lost Person Behavior}, beginning with searches near the LKP (ring 1) and then extending to ring 2 and beyond as time passes and early hypotheses are confirmed or disproven (P6).
After each cycle, commanders review debriefs to adjust the assumed LKP, re-rank PoIs, redefine area or linear-feature assignments, and reallocate teams. 
professionals across roles emphasized that maintaining shared situational awareness under these conditions—juggling evolving profiles, changing priorities, and scattered units—creates a heavy coordination burden for incident commanders, who must both keep the global search picture coherent and respond quickly to new information when updating the plan.
\noindent\textbf{Current tools and resources}
Across all five agencies, participants described having limited digital support for urban search operations. 
professionals emphasized that historical lost person data potentially valuable for analysis and predictive modeling, is difficult to access or aggregate (P6). 
In practice, they rely on a small set of geographic tools with narrow, task-specific utility rather than end-to-end support. 
SAR teams occasionally use SARTopo (P5) or CALTopo (P8) to switch between geographic layers, city agencies rely on ESRI-based tools (P4) to view maps, draw search areas, and display field searchers’ trajectories. 
Across agencies, the most consistent and widely used resource remains the \emph{Lost Person Behavior} book (P1, P2, P6), which provides conceptual guidance.
Surveillance cameras (e,g, Ring camera) are potentially to capture the lost person to show the clothing (P2, P4, P5). 

\subsection{Operational Challenges}
Although the five agencies differed substantially in staffing levels, specialized assets, and call volume, participants reported largely shared operational challenges in conducting urban lost-person searches.
Across agencies, professionals described markedly different baselines—ranging from resource-rich deployments (e.g., access to aviation assets such as helicopters) to understaffed teams managing high-frequency reports in dense urban environments.
For example, professionals from larger, well-resourced counties described the ability to escalate with specialized assets(e.g. drones, helicopters) when warranted (P8), whereas professionals from smaller or understaffed teams emphasized needing to “do more with less,” relying on limited personnel, time-compressed planning cycles, and rapidly changing field conditions (P1–P2, P5–P7).
Together, these contrasts highlight the importance of agency-adaptive support: future tools should accommodate heterogeneous resources while addressing common challenges shared across agencies.

\noindent\textbf{Partition the workforce across complex scenarios.}  
Across agencies, participants emphasized persistent understaffing as a structural constraint on search coverage. 
Incident commanders often struggle to assemble enough personnel to cover multiple zones, especially when the search area expands rapidly (P1, P6). 
professionals noted that limited manpower restricts how many PoIs or linear features can be searched concurrently, increasing the likelihood of gaps or delays.
As P1 summarized, ``even small reductions in workload matter,'' expressing openness to AI or robotic assistance if such tools could offset staffing shortages or reduce coordination overhead.

\noindent\textbf{Maintaining shared situational awareness during dynamic replanning is cognitively demanding.}  
Search operations evolve quickly as teams report clues, negative results, or updated sightings. Participants described a continual cycle of briefing, updating, and reprioritizing tasks, often under severe time pressure (P3, P6).
Commanders must revise search zones, reassign teams, and adjust assumptions about the LKP while keeping all units aligned. 
professionals reported that maintaining shared situational awareness during these rapid updates—especially across dispersed personnel—poses a significant coordination burden.

\noindent\textbf{Highly individualized profiles challenge standardized planning models.}  
While the ring model and profile categories offer useful structure, participants stressed that each missing person behaves differently in practice (P1–P6). 
Commanders must create a dynamically customized profile that incorporates idiosyncratic habits, personal history, and family-provided PoIs. 
Because these individualized patterns often diverge from the generalized behavioral templates, professionals rely heavily on ad-hoc reasoning and manual adjustments to create a plausible search plan. 
This limits the consistency and scalability of planning, making route design and prioritization difficult to generalize.

\noindent\textbf{High physical and cognitive demands on field searchers.}  
Field professionals described the dual burden of physically covering terrain while maintaining attention to subtle cues, potential hazards, and moving targets. 
Missing persons may continue to walk or change direction, requiring teams to re-scan areas or revisit sectors (P2, P4, P5). 
Meanwhile, incident commanders face heavy cognitive load as they attempt to make timely decisions, integrate new information, and direct teams effectively. 
Participants noted that this combination of physical effort and cognitive vigilance increases fatigue and raises the risk of missed detections.

\section{Discussion}
\begin{table*}[t]
\centering
\vspace{-0.8em}
\small
\setlength{\tabcolsep}{3.5pt}
\renewcommand{\arraystretch}{1.12}
\resizebox{\textwidth}{!}{%
\begin{tabularx}{\textwidth}{p{0.14\textwidth} X p{0.24\textwidth} p{0.135\textwidth}}
\toprule
\textbf{Results Theme (RQ)} & \textbf{Key Empirical Finding (Sec.4)} & \textbf{Implication (Sec.5)} & \textbf{DRs} \\
\midrule
\textbf{Profile-driven planning (RQ1)} &
Profiles shape search plans through a combination of general person-type behavior patterns and individualized factors (e.g., PoIs/PoDs and situational context), while witness-reported appearance colors are often unreliable.&
Make the profile a living artifact; support evidence-backed cueing without rigid templates. &
DR1, DR4, DR5 \\
\midrule
\textbf{Coordination in planning (RQ1)} &
Commanders plan and re-plan via point/area/linear modes through brief–debrief cycles, synthesizing fragmented info and deploying limited personnel under high SA burden.&
Support fast, justifiable replanning; reduce coordination SA cost by LLM and map-centric summary. &
DR1, DR3, DR6 \\
\midrule
\textbf{Tools/resources (RQ1)} &
Agencies stitch narrow tools (GIS, SAR maps, LPB book) with limited end-to-end support. &
Unify artifacts, assignments, evidence, and replanning history in one operational surface. &
DR3, DR6 \\
\midrule
\textbf{Understaffing (RQ1)} &
Limited manpower constrains concurrent coverage and increases gaps/delays. &
UGVs absorb repeatable coverage/re-coverage with low added friction. &
DR2, DR3 \\
\midrule
\textbf{Dynamic replanning burden (RQ1/RQ3)} &
Rapid updates from negative results/sightings create heavy cognitive load. &
AI should enable notice--communicate--replan with explicit rationale and uncertainty. &
DR1, DR3, DR6 \\
\midrule
\textbf{Individualized profiles (RQ1/RQ3)} &
Case-by-case behavior diverges from general templates; planning relies on ad-hoc adjustments. &
Provide flexible overrides and editable representations aligned with operational logic. &
DR1, DR4 \\
\midrule
\textbf{Field load (RQ1/RQ3)} &
Fatigue and vigilance demands raise the missed-detection risk. &
Robots as mobile sensing; UI lead with guardrails to avoid overload.&
DR2, DR5 \\
\midrule
\textbf{UGV roles \& limits (RQ2)} &
UGVs fit repeatable patrol/coverage (sidewalks/water edges); adoption requires low-burden operation and clear boundaries. &
Frame as complements; prioritize lightweight supervision and accountability. &
DR2, DR3, DR6 \\
\bottomrule
\end{tabularx}}
\vspace{-0.4em}
\caption{Crosswalk from Results to Discussion and distilled Design Requirements (DR1--DR6).}
\vspace{-0.8em}
\label{tab:rq-results-discussion-dr}
\end{table*}
Our focus groups surfaced a number of challenges faced by police in urban search.
These challenges were chosen in part because they represent areas where existing research has not focused much attention, but where progress may have a strong impact for professionals. 
Our study was motivated by three research questions: understanding real-world urban search practice and constraints (RQ1), identifying meaningful roles and limits for affordable ground-robot fleets (RQ2), and distilling design opportunities for human-centered, accountable tools that combine robots, computer vision, and AI-assisted reasoning (RQ3).
Across five Virginia agencies, participants described a practice that is both distributed and iterative: commanders continuously partition limited personnel across evolving hypotheses, revise areas and linear features as new brief/debrief updates arrive, and rely on external artifacts (maps, templates, ad-hoc book) to maintain shared awareness under time pressure.
These realities explain why purely ``optimal'' routing or single-shot prediction is rarely actionable; instead, what practitioners need are systems that (1) scale coverage without increasing coordination overhead, (2) update routes while preserving justification, and (3) integrate person-specific cues and field evidence without forcing rigid templates.
Below, we translate our findings into implications for deployable robot-fleet support.

\subsection{Optimized Robot Fleets as Complements}
\label{sec:optimized_robot_fleet}
Our findings suggest that affordable UGV fleets are most valuable when positioned as a \emph{complement} to both aerial assets (when available) and human teams, not as a replacement. 
In everyday urban search, the highest-leverage role for ground robots is to absorb repetitive, time-consuming coverage and re-coverage that remains operationally necessary (e.g., predefined sweeps, re-scanning high-probability areas, and patrolling linear features such as sidewalks, water edges, or trail segments). 
By taking on these “repeatable” tasks, UGVs can free officers to focus on judgment-intensive work that cannot be delegated: interviewing families, coordinating units, negotiating access, and managing community interactions.

Crucially, this complement framing clarifies what a robot fleet should optimize for coverage with low coordination cost. 
Robots only reduce workload if they do not introduce new operational friction. 
This motivates an explicit design principle: \emph{minimize operational cost without adding burden}. 
Affordable fleets must be realistic for small and mid-sized agencies, which means affordability beyond purchase price including easy maintenance, rapid deployment, and low downtime in routine field conditions. 
In parallel, operation should require no technical expertise: commanders should be able to assign tasks, adjust priorities, and trigger re-scans through simple interactions (e.g., selecting areas/paths on a map or issuing short natural-language commands), while the system handles low-level control, multi-robot coordination, and recovery from common failures.

To function as practical force multipliers, UGV fleets should therefore behave as collaborative, autonomous teammates rather than teleoperated devices that demand constant attention. 
In practice, a robots fleet should execute assigned routes with lightweight supervision and continuously report status in a way that fits commanders’ iterative replanning loop. 
When robots are low-cost, low-maintenance, and low-burden to operate, they can scale search coverage while preserving human authority and reducing cognitive load, which is a prerequisite for adoption in real-world urban search operations.

\subsection{Optimized AI Support for Urban Search}
\label{sec:optimized_ai_support}

Beyond expanding coverage with robot fleets, agencies need decision support that helps them \emph{notice, communicate, and re-plan} under time pressure and fatigue. RQ2 and RQ3 highlight a consistent theme: professionals are open to AI assistance, but only when it aligns with the operational logic they already use. In practice, commanders translate profiles into interlocking planning modes---point-based (PoIs/PoDs), area-based (rings and zones), and linear-feature–based (sidewalks, creeks, water edges)---starting from the LKP and direction of travel, then repeatedly updating priorities as negative results and new sightings arrive. In this dynamic replanning loop, “smart” tools are most useful not when they produce a single optimal plan, but when they support fast, justifiable updates that keep all units aligned.

\textbf{Perception support.}
Participants emphasized that search decisions often hinge on visual and contextual cues (recent clothing, belongings, posture, and surveillance clips), yet descriptions can be inconsistent or ambiguous (e.g., color naming differences or incomplete reports). AI can help by structuring and surfacing \emph{evidence} from heterogeneous sources---robot video, fixed cameras when available, and field updates: indexing short snippets around likely locations and linear features, enabling attribute- or appearance-based retrieval, and highlighting candidate cues for rapid human validation. Here, computer-vision capabilities need to be carefully scoped: while broad face-recognition raises legal and community concerns, appearance-based retrieval and reacquisition for a \emph{specific, authorized missing-person case} can provide a practical way to re-find individuals along routes or in crowded scenes. Across these capabilities, outputs should be framed as \emph{actionable leads} rather than definitive conclusions, with lightweight controls (adjustable sensitivity, clear dismissal options) to prevent alert fatigue and overload.

\textbf{Planning support (adaptive updates with justification).}
Professionals described urban search as a continuous, profile-driven replanning process rather than a one-shot optimization problem. AI assistance can reduce coordination burden by translating new clues, negative results, and evolving hypotheses into operationally meaningful revisions---re-ranking PoIs, suggesting targeted re-scans of specific zones or linear features, and proposing resource reallocations---while keeping the “why” explicit. Each suggested change should clearly indicate (1) what changed, (2) what triggered the change (with links back to the underlying evidence), and (3) what uncertainty remains, so commanders can accept, edit, or reject recommendations while preserving shared situational awareness. To remain trustworthy, any AI-assisted routing component should produce not only a route but also a traceable rationale tied to familiar constructs (LPB templates, rings, PoIs/PoDs, linear features, and LKP updates) and preserve an audit trail of changes. Rather than treating the profile as static input, future systems should expose an operator-facing profile panel that is compact, editable, and actionable, and treat profile information as a living artifact that evolves through debriefs and evidence, driving real-time replanning and re-ranking while preserving interpretability and accountability.

\subsection{Police-Centered Operational Interface}
\label{sec:police_centered_ui}

Our data underscores that adoption hinges on interface design that \emph{fits police workflows}. Commanders already reason in map-centric terms---rings, zones, PoIs/PoDs, and linear features---but lack integrated decision-making tools for urban search. In our focus groups, agencies described stitching together narrow tools (e.g., SARTopo, municipal GIS, ESRI-based maps) and paper references (e.g., the Lost Person Behavior book) rather than using a unified system. In this context, the goal is not simply to “add robots” or “add AI,” but to design an operational interface that brings these capabilities together and makes coordination easier, faster, and more defensible.

First, the interface should support \emph{multi-level externalization} and \emph{low-friction tasking}. Incident commanders need to see what has been searched, what remains, and why priorities changed, without reconstructing state from scattered radio logs and mental notes. A single operational surface should show zones, PoIs/PoDs, linear features, assigned human teams and robots, AI-suggested updates, and replanning history, all directly manipulable with minimal interaction. Our findings motivate two complementary interaction styles: natural language commands (e.g., “re-scan this sidewalk segment” or “send one unit to PoD B”) and direct map interactions (drawing or selecting routes, areas, or points). Both should be paired with one-click operations for common actions such as re-scanning an area, covering a linear feature, or patrolling a PoD. This hybrid approach matches how professionals already think, and allows human teams and robots to be tasked through the same interaction grammar, with the system handling low-level scheduling and coordination.

Second, police-centered interface design must explicitly address \emph{accountability} and frame perception support as \emph{cueing and communication augmentation} rather than surveillance. Urban searches must be defensible, not only effective: commanders need to explain why an area was prioritized, why a route changed, and how resources were allocated over time. A practical implication is to treat replanning as a first-class event: every AI-supported update or robot reassignment should record what changed, which evidence triggered the change, and when and by whom it was approved, making it easier to review and, if needed, challenge decisions later. At the same time, field searchers face a dual burden of physically covering terrain while staying alert for subtle cues and hazards, with fatigue increasing the risk of missed detections. Robots can act as mobile perception platforms that stream video and capture keyframes, while AI components help surface candidate cues and summarize debriefs; yet the interface must allow professionals to control when and how these cues are surfaced—through filters, priorities, and simple acknowledgment or dismissal—to avoid overload and preserve trust. In combination, a police-centered operational interface, optimized robot fleets, and optimized AI support form an integrated, human-led search system rather than a collection of disconnected tools.

\subsection{Design Requirements Summary}
\label{sec:design_requirements_summary}
Synthesizing across current practice and operational challenges, we distill the following design requirements for future urban search systems that integrate multi-robot fleets, computer vision, and AI-assisted reasoning:

\noindent\textbf{DR1: Real-time adaptive route planning grounded in operational logic.}
Systems should generate and update routes based on LPB guidance, LKP and direction of travel, PoIs/PoDs, environmental context, and planning modes that match existing practice (point-based, area-based, and linear-feature–based).
Updates should be triggered by debriefs, negative results, and sightings, and accompanied by clear justifications that preserve shared situational awareness (Practices: coordinated planning and replanning; Challenges: dynamic replanning and coordination burden).

\noindent\textbf{DR2: Multi-robot task assignment and autonomous execution with lightweight supervision.}
Systems should enable commanders to assign robots to cover zones, patrol linear features, re-scan areas, and monitor PoDs, while minimizing the need for continuous teleoperation.
Robots should report status and completion in ways that support the commanders’ iterative replanning loop (Practices: workforce partitioning and zone assignment; Challenges: understaffing and cognitive load).

\noindent\textbf{DR3: Police-centered planning and control interface with low-friction tasking.}
The interface should allow operators to edit routes and assignments through direct map interaction (draw/select areas, routes, and points), and optionally through concise natural-language commands for common actions.
It should unify human-team and robot tasking into the same interaction grammar to reduce coordination overhead (Practices: map-centric reasoning; Challenges: coordination burden).

\noindent\textbf{DR4: Operator-facing profile representation as a living artifact.}
Systems should provide a compact, easily editable profile panel that captures key attributes (age, condition, non-verbal status), likely mobility patterns, time-of-day expectations, prioritized PoIs, and reliable appearance evidence.
The profile should be continuously revisable as new evidence arrives, enabling case-specific planning without forcing rigid templates (Practices: profile-driven planning; Challenges: individualized profiles).

\noindent\textbf{DR5: Vision-based cueing, retrieval, and evidence management with guardrails.}
Systems should reconcile inconsistent human-provided appearance descriptions (e.g., color ambiguity) using available imagery (photos/surveillance/robot video), and support appearance-based retrieval and reacquisition for authorized cases.
Outputs should be framed as leads with adjustable sensitivity and clear dismissal controls to prevent overload and preserve trust (Practices: reliance on photos/surveillance; Challenges: ambiguity and cognitive strain).

\noindent\textbf{DR6: Replanning accountability and auditability.}
Systems should maintain a traceable record of replanning and reassignment decisions, linking changes to triggers (field updates, evidence cues) and capturing when and by whom updates were approved.
This audit trail supports defensibility and helps maintain coherent shared situational awareness across distributed teams (Practices: iterative briefing/debriefing; Challenges: dynamic replanning under pressure).

\subsection{limitations}
Our findings should be interpreted with several limitations.
First, our focus groups included eight officers across five Virginia law enforcement agencies; while participants spanned roles from detectives to incident commanders, the sample is region- and context-specific
Second, because ground robot fleets are not yet widely deployed in routine urban search operations, we relied on focus groups to probe realistic hypothetical integration scenarios rather than observing long-term in-the-wild use.
Third, we did not audio-record sessions to respect confidentiality and departmental protocols; we mitigated this through three-researcher collaborative note-taking and immediate consolidation, but this may miss nuances compared to full transcripts.

\section{Conclusion}
We investigated how public safety professionals conduct urban lost-person searches and how ground-robot fleets might be integrated into these workflows.
Guided by three research questions, we conducted focus group interviews with eight police officers across five Virginia agencies to surface current practices, practical constraints, and adoption-relevant opportunities for robot fleets.
Our findings show that practitioners face persistent understaffing, high coordination burden during dynamic replanning, difficulty operationalizing highly individualized profiles, and substantial physical and cognitive strain in the field.
In response, we identify design opportunities for scalable multi-robot support, adaptive and explainable route planning grounded in real search logic, profile-sensitive systems that ingest contextual cues and evidence, and vision-assisted cueing that reduces fatigue while preserving trust.
We further distill actionable design requirements which including real-time adaptive route planning, multi-robot task assignment, police-centered route control interfaces, operator-facing profile panels, and vision-based detection/retrieval with appropriate guardrails that can guide researchers across HCI, robotics, and computer vision in developing deployable, accountable, and human-centered urban search support systems.

\bibliographystyle{abbrv}
\bibliography{robotSAR}

@article{koester2008lost,
  title={Lost person behavior: A search and rescue},
  author={Koester, Robert J},
  journal={dbs Productions LLC},
  year={2008}
}

@article{pub1991national,
  title={National Search and Rescue Manual Volume i: National Search and Rescue System},
  author={PUB, JOINT},
  year={1991}
}

@article{wollan2004incorporating,
  title={Incorporating heuristically generated search patterns in search and rescue},
  author={Wollan, Helen},
  journal={University of Edinburgh},
  year={2004}
}

@inproceedings{xu2024reinforcement,
  title={Reinforcement learning for wheeled mobility on vertically challenging terrain},
  author={Xu, Tong and Pan, Chenhui and Xiao, Xuesu},
  booktitle={2024 IEEE International Symposium on Safety Security Rescue Robotics (SSRR)},
  pages={125--130},
  year={2024},
  organization={IEEE}
}

@inproceedings{du2022unknown,
  title={Unknown-aware object detection: Learning what you don't know from videos in the wild},
  author={Du, Xuefeng and Wang, Xin and Gozum, Gabriel and Li, Yixuan},
  booktitle={Proceedings of the IEEE/CVF conference on computer vision and pattern recognition},
  pages={13678--13688},
  year={2022}
}

@inproceedings{jiang2023cross,
  title={Cross-modal implicit relation reasoning and aligning for text-to-image person retrieval},
  author={Jiang, Ding and Ye, Mang},
  booktitle={Proceedings of the IEEE/CVF conference on computer vision and pattern recognition},
  pages={2787--2797},
  year={2023}
}

@article{cao2025exploring,
  title={Exploring LLM-based multi-agent situation awareness for zero-trust space-air-ground integrated network},
  author={Cao, Xinye and Nan, Guoshun and Guo, Hongcan and Mu, Hanqing and Wang, Long and Lin, Yihan and Zhou, Qinchuan and Li, Jiayi and Qin, Baohua and Cui, Qimei and others},
  journal={IEEE Journal on Selected Areas in Communications},
  year={2025},
  publisher={IEEE}
}

@article{xiao2022motion,
  title={Motion planning and control for mobile robot navigation using machine learning: a survey},
  author={Xiao, Xuesu and Liu, Bo and Warnell, Garrett and Stone, Peter},
  journal={Autonomous Robots},
  volume={46},
  number={5},
  pages={569--597},
  year={2022},
  publisher={Springer}
}

@article{burgard2005coordinated,
  title={Coordinated multi-robot exploration},
  author={Burgard, Wolfram and Moors, Mark and Stachniss, Cyrill and Schneider, Frank E},
  journal={IEEE Transactions on robotics},
  volume={21},
  number={3},
  pages={376--386},
  year={2005},
  publisher={IEEE}
}

@INPROCEEDINGS{11247499,
  author={Berneburg, James and Wang, Xuan and Xiao, Xuesu and Shishika, Daigo},
  booktitle={2025 IEEE/RSJ International Conference on Intelligent Robots and Systems (IROS)}, 
  title={Multi-Robot Coordination in an Adversarial Graph-Traversal Game}, 
  year={2025},
  volume={},
  number={},
  pages={3915-3922},
  doi={10.1109/IROS60139.2025.11247499}}

@INPROCEEDINGS{11247618,
  author={Payandeh, Amirreza and Song, Daeun and Nazeri, Mohammad and Liang, Jing and Mukherjee, Praneel and Raj, Amir Hossain and Kong, Yangzhe and Manocha, Dinesh and Xiao, Xuesu},
  booktitle={2025 IEEE/RSJ International Conference on Intelligent Robots and Systems (IROS)}, 
  title={Social-LLaVA: Enhancing Social Robot Navigation through Human-Language Reasoning}, 
  year={2025},
  volume={},
  number={},
  pages={17192-17198},
  doi={10.1109/IROS60139.2025.11247618}}

@article{hashimoto2022agent,
  title={An agent-based model reveals lost person behavior based on data from wilderness search and rescue},
  author={Hashimoto, Amanda and Heintzman, Larkin and Koester, Robert and Abaid, Nicole},
  journal={Scientific reports},
  volume={12},
  number={1},
  pages={5873},
  year={2022},
  publisher={Nature Publishing Group UK London}
}

@book{stone2007theory,
  author    = {Stone, Lawrence D.},
  title     = {Theory of Optimal Search},
  year      = {2007},
  publisher = {INFORMS},
  address   = {Catonsville, MD},
  edition   = {2nd},
  isbn      = {978-0-9843378-0-8}
}

@misc{koester2010international,
  title={International Search and Rescue Incident Database (ISRID)},
  author={Koester, Robert J},
  year={2010}
}

@inproceedings{nguyen2023finding,
  title={Finding a Needle in the Haystack: Predicting the Location of Lost People Using Agent-Based Modeling and Behavioral Inertia},
  author={Nguyen, John and Joseph, Caroline and Richardson, Bailey and Hayes, Roy and Pakula, Ricardo and Koester, Robert},
  booktitle={2023 Systems and Information Engineering Design Symposium (SIEDS)},
  pages={78--83},
  year={2023},
  organization={IEEE}
}

@article{ewers2025predictive,
  title={Predictive Probability Density Mapping for Search and Rescue Using An Agent-Based Approach with Sparse Data},
  author={Ewers, Jan-Hendrik and Anderson, David and Thomson, Douglas},
  journal={IEEE Access},
  year={2025},
  publisher={IEEE}
}

@article{ewers2023optimal,
  title={Optimal path planning using psychological profiling in drone-assisted missing person search},
  author={Ewers, Jan-Hendrik and Anderson, David and Thomson, Douglas},
  journal={Advanced Control for Applications: Engineering and Industrial Systems},
  volume={5},
  number={4},
  pages={e167},
  year={2023},
  publisher={Wiley Online Library}
}

@article{sava2016evaluating,
  title={Evaluating lost person behavior models},
  author={Sava, Elena and Twardy, Charles and Koester, Robert and Sonwalkar, Mukul},
  journal={Transactions in GIS},
  volume={20},
  number={1},
  pages={38--53},
  year={2016},
  publisher={Wiley Online Library}
}

@article{cohen2015investigation,
  title={An investigation of operational decision making in situ: Incident command in the UK fire and rescue service},
  author={Cohen-Hatton, Sabrina R and Butler, Philip C and Honey, Robert C},
  journal={Human Factors},
  volume={57},
  number={5},
  pages={793--804},
  year={2015},
  publisher={Sage Publications Sage CA: Los Angeles, CA}
}

@inproceedings{alharthi2021activity,
  title={An activity theory analysis of search \& rescue collective sensemaking and planning practices},
  author={Alharthi, Sultan A and LaLone, Nicolas James and Sharma, Hitesh Nidhi and Dolgov, Igor and Toups Dugas, Phoebe O},
  booktitle={Proceedings of the 2021 CHI Conference on Human Factors in Computing Systems},
  pages={1--20},
  year={2021}
}

@article{ferguson2023training,
  title={Training police search and rescue teams: Implications for missing persons work},
  author={Ferguson, Lorna and Gaub, Janne E},
  journal={Criminology \& criminal justice},
  volume={23},
  number={3},
  pages={431--449},
  year={2023},
  publisher={SAGE Publications Sage UK: London, England}
}

@article{hill2012cognition,
  title={Cognition in the woods: Biases in probability judgments by search and rescue planners},
  author={Hill, Kenneth A},
  journal={Judgment and Decision Making},
  volume={7},
  number={4},
  pages={488--498},
  year={2012},
  publisher={Cambridge University Press}
}

@article{doherty2014analysis,
  title={An analysis of probability of area techniques for missing persons in Yosemite National Park},
  author={Doherty, Paul J and Guo, Quinghua and Doke, Jared and Ferguson, Don},
  journal={Applied Geography},
  volume={47},
  pages={99--110},
  year={2014},
  publisher={Elsevier}
}

@article{prati2010self,
  title={Self-efficacy moderates the relationship between stress appraisal and quality of life among rescue workers},
  author={Prati, Gabriele and Pietrantoni, Luca and Cicognani, Elvira},
  journal={Anxiety, Stress, \& Coping},
  volume={23},
  number={4},
  pages={463--470},
  year={2010},
  publisher={Taylor \& Francis}
}

@article{adams2007search,
  title={Search is a time-critical event: when search and rescue missions may become futile},
  author={Adams, Annette L and Schmidt, Terri A and Newgard, Craig D and Federiuk, Carol S and Christie, Michael and Scorvo, Sean and DeFreest, Melissa},
  journal={Wilderness \& environmental medicine},
  volume={18},
  number={2},
  pages={95--101},
  year={2007},
  publisher={SAGE Publications Sage CA: Los Angeles, CA}
}

@inproceedings{rekleitis2002multi,
  title={Multi-robot cooperative localization: a study of trade-offs between efficiency and accuracy},
  author={Rekleitis, Ioannis M and Dudek, Gregory and Milios, Evangelos E},
  booktitle={IEEE/RSJ International Conference on Intelligent Robots and Systems},
  volume={3},
  pages={2690--2695},
  year={2002},
  organization={IEEE}
}

@article{meng2024llm,
  title={Llm-a*: Large language model enhanced incremental heuristic search on path planning},
  author={Meng, Silin and Wang, Yiwei and Yang, Cheng-Fu and Peng, Nanyun and Chang, Kai-Wei},
  journal={arXiv preprint arXiv:2407.02511},
  year={2024}
}

@inproceedings{zhou2024hugs,
  title={Hugs: Holistic urban 3d scene understanding via gaussian splatting},
  author={Zhou, Hongyu and Shao, Jiahao and Xu, Lu and Bai, Dongfeng and Qiu, Weichao and Liu, Bingbing and Wang, Yue and Geiger, Andreas and Liao, Yiyi},
  booktitle={Proceedings of the IEEE/CVF Conference on Computer Vision and Pattern Recognition},
  pages={21336--21345},
  year={2024}
}

@article{alotaibi2019lsar,
  title={Lsar: Multi-uav collaboration for search and rescue missions},
  author={Alotaibi, Ebtehal Turki and Alqefari, Shahad Saleh and Koubaa, Anis},
  journal={Ieee Access},
  volume={7},
  pages={55817--55832},
  year={2019},
  publisher={IEEE}
}

@article{vinnikov2022occupational,
  title={Occupational burn-out, fatigue and stress in professional rescuers: a cross-sectional study in Kazakhstan},
  author={Vinnikov, Denis and Kapanova, Gulnara and Romanova, Zhanna and Krugovykh, Ilya and Kalmakhanov, Sundetgali and Ualiyeva, Aliya and Baigonova, Kaini and Tulekov, Zhangir and Ongarbaeva, Damet},
  journal={BMJ open},
  volume={12},
  number={6},
  pages={e057935},
  year={2022},
  publisher={British Medical Journal Publishing Group}
}

@book{young2007urban,
  title={Urban search: Managing missing person searches in the urban environment},
  author={Young, Christopher S and Wehbring, John},
  year={2007},
  publisher={DbS Productions}
}

@article{feinberg2021reconceptualizing,
  title={Reconceptualizing rapid responses as a speededness indicator in high-stakes assessments},
  author={Feinberg, Richard and Jurich, Daniel and Wise, Steven L},
  journal={Applied Measurement in Education},
  volume={34},
  number={4},
  pages={312--326},
  year={2021},
  publisher={Taylor \& Francis}
}

@article{jiang2024urbanllm,
  title={Urbanllm: Autonomous urban activity planning and management with large language models},
  author={Jiang, Yue and Chao, Qin and Chen, Yile and Li, Xiucheng and Liu, Shuai and Cong, Gao},
  journal={arXiv preprint arXiv:2406.12360},
  year={2024}
}

@article{violakis2025leveraging,
  title={Leveraging large language models for enhanced simulation-based learning in police and law enforcement},
  author={Violakis, Petros},
  journal={Policing: A Journal of Policy and Practice},
  volume={19},
  pages={paaf012},
  year={2025},
  publisher={Oxford University Press UK}
}

@inproceedings{Roberts2016UnmannedVC,
  title={Unmanned Vehicle Collaboration Research Environment for Maritime Search and Rescue},
  author={William Roberts and Kelly Griendling and A. Gray and Dimitri N. Mavris},
  year={2016},
  url={https://api.semanticscholar.org/CorpusID:231994310}
}

@inproceedings{Grayson2014SearchR,
  title={Search \& Rescue using Multi-Robot Systems},
  author={Siobh{\'a}n Grayson},
  year={2014},
  url={https://api.semanticscholar.org/CorpusID:33345160}
}

@article{ko2009intelligent,
  title={Intelligent robot-assisted humanitarian search and rescue system},
  author={Ko, Albert WY and Lau, Henry YK},
  journal={International Journal of Advanced Robotic Systems},
  volume={6},
  number={2},
  pages={12},
  year={2009},
  publisher={SAGE Publications Sage UK: London, England}
}

@article{casper2003human,
  title={Human-robot interactions during the robot-assisted urban search and rescue response at the world trade center},
  author={Casper, Jennifer and Murphy, Robin R.},
  journal={IEEE Transactions on Systems, Man, and Cybernetics, Part B (Cybernetics)},
  volume={33},
  number={3},
  pages={367--385},
  year={2003},
  publisher={IEEE}
}

@article{quero2025unmanned,
  title={Unmanned aerial systems in search and rescue: A global perspective on current challenges and future applications},
  author={Quero, Carlos Osorio and Martinez-Carranza, Jose},
  journal={International Journal of Disaster Risk Reduction},
  pages={105199},
  year={2025},
  publisher={Elsevier}
}

@article{kashino2020aerial,
  title={Aerial wilderness search and rescue with ground support},
  author={Kashino, Zendai and Nejat, Goldie and Benhabib, Beno},
  journal={Journal of Intelligent \& Robotic Systems},
  volume={99},
  number={1},
  pages={147--163},
  year={2020},
  publisher={Springer}
}

@article{hong2019investigating,
  title     = {Investigating Human-Robot Teams for Learning-Based Semi-autonomous Control in Urban Search and Rescue Environments},
  author    = {Hong, A. and Igharoro, O. and Liu, Yugang and Niroui, Farzad and Nejat, Goldie and Benhabib, Beno},
  journal   = {Journal of Intelligent \& Robotic Systems},
  volume    = {94},
  number    = {3},
  pages     = {669--686},
  year      = {2019},
  publisher = {Springer},
  doi       = {10.1007/s10846-018-0899-0}
}

@inproceedings{glantz2020uav,
  title        = {UAV Use in Disaster Management},
  author       = {Glantz, Edward J. and Ritter, Frank E. and Gilbreath, Don and Stager, Sarah J. and Anton, Alexandra and Emani, Rahul},
  booktitle    = {ISCRAM 2020 - Proceedings of the 17th International Conference on Information Systems for Crisis Response and Management},
  pages        = {914--921},
  year         = {2020},
  editor       = {Hughes, Amanda Lee and McNeill, Fiona and Zobel, Christopher W.},
  organization = {Information Systems for Crisis Response and Management, ISCRAM},
  address      = {Blacksburg, VA}
}

@article{doroodgar2014learning,
  title={A learning-based semi-autonomous controller for robotic exploration of unknown disaster scenes while searching for victims},
  author={Doroodgar, Barzin and Liu, Yugang and Nejat, Goldie},
  journal={IEEE Transactions on Cybernetics},
  volume={44},
  number={12},
  pages={2719--2732},
  year={2014},
  publisher={IEEE}
}

@article{burke2004moonlight,
  title={Moonlight in Miami: Field study of human-robot interaction in the context of an urban search and rescue disaster response training exercise},
  author={Burke, Jennifer L and Murphy, Robin R and Coovert, Michael D and Riddle, Dawn L},
  journal={Human--Computer Interaction},
  volume={19},
  number={1-2},
  pages={85--116},
  year={2004},
  publisher={Taylor \& Francis}
}

@article{nytimes2023K5,
  author = {Cramer, Maria and Rubinstein, Dana and Meko, Hurubie},
  title = {400-Pound N.Y.P.D. Robot Gets Tryout in Times Square Subway Station},
  journal = {The New York Times},
  date = {2023-09-22},
  url = {https://www.nytimes.com/2023/09/22/nyregion/police-robot-times-square-nyc.html}
}

@inproceedings{williams2020collaborative,
  title={Collaborative multi-robot multi-human teams in search and rescue.},
  author={Williams, R},
  booktitle={Proceedings of the international ISCRAM conference},
  volume={17},
  year={2020}
}

@article{francos2023role,
  title={On the role and opportunities in teamwork design for advanced multi-robot search systems},
  author={Francos, Roee M and Bruckstein, Alfred M},
  journal={Frontiers in Robotics and AI},
  volume={10},
  pages={1089062},
  year={2023},
  publisher={Frontiers Media SA}
}

@article{barisic2025unlocking,
  title={Unlocking Thermal Aerial Imaging: Synthetic Enhancement of UAV Datasets},
  author={Barisic Kulas, Antonella and Brkljac, Nikola},
  journal={arXiv preprint arXiv:2507.06797},
  year={2025}
}

@article{dhiman2022firefighting,
  title={Firefighting robot with deep learning and machine vision},
  author={Dhiman, Amit and Shah, Neel and Adhikari, Pranali and Kumbhar, Sayali and Dhanjal, Inderjit Singh and Mehendale, Ninad},
  journal={Neural Computing and Applications},
  volume={34},
  number={4},
  pages={2831--2839},
  year={2022},
  publisher={Springer}
}

@inproceedings{guo2020motion_blur_dnn_neurips,
  title     = {Watch out! Motion is Blurring the Vision of Your Deep Neural Networks},
  author    = {Guo, Qing and Juefei-Xu, Felix and Xie, Xiaofei and Ma, Lei and Wang, Jian and Yu, Bing and Feng, Wei and Liu, Yang},
  booktitle = {Advances in Neural Information Processing Systems},
  volume    = {33},
  pages     = {1--11},
  year      = {2020}
}

@inproceedings{li2024llm_poi,
  title     = {Large Language Models for Next Point-of-Interest Recommendation},
  author    = {Li, Peibo and de Rijke, Maarten and Xue, Hao and Ao, Shuang and Song, Yang and Salim, Flora D.},
  booktitle = {Proceedings of the 47th International ACM SIGIR Conference on Research and Development in Information Retrieval (SIGIR '24)},
  year      = {2024},
  doi       = {10.1145/3626772.3657840},
}

@inproceedings{bucinca2021trust,
  author    = {Bu{\c{c}}inca, Zana and Malaya, Maja Barbara and Gajos, Krzysztof Z.},
  title     = {To Trust or to Think: Cognitive Forcing Functions Can Reduce Overreliance on AI in AI-assisted Decision-making},
  booktitle = {Proceedings of the 2021 CHI Conference on Human Factors in Computing Systems (CHI '21)},
  year      = {2021},
  doi       = {10.1145/3449287}
}

@inproceedings{spatharioti2025llmsearch,
  author  = {Spatharioti, Sofia and Rothschild, Daniel and O'Hara, Kieron and others},
  title   = {Effects of LLM-based Search on Decision Making: Speed, Accuracy, and Overreliance},
  booktitle = {Proceedings of the 2025 CHI Conference on Human Factors in Computing Systems (CHI '25)},
  year    = {2025},
  doi     = {10.1145/3706599.3720581}
}

@inproceedings{karusala2024publicservices,
  author    = {Karusala, Naveena and others},
  title     = {Enabling Contestability in Public Services: Understanding Workflows, Constraints, and Opportunities},
  booktitle = {Proceedings of the 2024 CHI Conference on Human Factors in Computing Systems (CHI '24)},
  year      = {2024},
  doi       = {10.1145/3706599.3721096}
}

@inproceedings{herrewijnen2024xailaw,
  author    = {Herrewijnen, Jan and others},
  title     = {Requirements and Attitudes towards Explainable AI in Law Enforcement Decision-making},
  booktitle = {Proceedings of the 2024 ACM Designing Interactive Systems Conference (DIS '24)},
  year      = {2024},
  doi       = {10.1145/3643834.3661629}
}

@misc{otal2024crisis,
  title        = {LLM-Assisted Crisis Management: Building Advanced LLM Platforms for Effective Emergency Response and Public Collaboration},
  author       = {Otal, Hakan T. and Canbaz, M. Abdullah},
  year         = {2024},
  eprint       = {2402.10908},
  archivePrefix= {arXiv},
  primaryClass = {cs.CL},
}

@misc{xiao2025llmadvisorllmbenchmarkcostefficient,
      title={LLM-Advisor: An LLM Benchmark for Cost-efficient Path Planning across Multiple Terrains}, 
      author={Ling Xiao and Toshihiko Yamasaki},
      year={2025},
      eprint={2503.01236},
      archivePrefix={arXiv},
      primaryClass={cs.RO},
      url={https://arxiv.org/abs/2503.01236}, 
}

@article{chen2024ekell,
  title   = {Efficient knowledge-enhanced large language model for emergency decision-making},
  author  = {Chen, Yuchen and others},
  journal = {International Journal of Disaster Risk Reduction},
  year    = {2024},
  volume  = {113},
  pages   = {104804},
  doi     = {10.1016/j.ijdrr.2024.104804},
}

@ARTICLE{8970561,
  author={Zeng, Zelong and Wang, Zhixiang and Wang, Zheng and Zheng, Yinqiang and Chuang, Yung-Yu and Satoh, Shin’ichi},
  journal={IEEE Transactions on Multimedia}, 
  title={Illumination-Adaptive Person Re-Identification}, 
  year={2020},
  volume={22},
  number={12},
  pages={3064-3074},
  doi={10.1109/TMM.2020.2969782}}

@article{queralta2020collaborative,
  title={Collaborative multi-robot search and rescue: Planning, coordination, perception, and active vision},
  author={Queralta, Jorge Pena and Taipalmaa, Jussi and Pullinen, Bilge Can and Sarker, Victor Kathan and Gia, Tuan Nguyen and Tenhunen, Hannu and Gabbouj, Moncef and Raitoharju, Jenni and Westerlund, Tomi},
  journal={Ieee Access},
  volume={8},
  pages={191617--191643},
  year={2020},
  publisher={IEEE}
}
\end{document}